\documentclass[aps,prb,reprint,superscriptaddress,nofootinbib,floatfix]{revtex4-2}
\usepackage[T1]{fontenc}
\usepackage{lmodern}
\usepackage{amsmath,amssymb,bm}
\usepackage{booktabs}
\usepackage{graphicx}
\usepackage{xcolor}
\usepackage{tikz}

\usetikzlibrary{arrows.meta,decorations.pathreplacing}
\newcounter{algorithm}
\usepackage[colorlinks=true,allcolors=blue]{hyperref}
\usepackage{orcidlink}

\begin{document}

\title{Exact environment checkpointing for memory-limited DMRG and TDVP}

\author{Nihal Krishna Shrivastava\,\orcidlink{0009-0000-7969-4217}}
\email{f20230986@goa.bits-pilani.ac.in}
\affiliation{Birla Institute of Technology and Science, Pilani, K K Birla Goa Campus, Goa 403726, India}
\author{Rudranil Basu\,\orcidlink{0000-0003-0655-0890}}
\email{rudranilb@goa.bits-pilani.ac.in}
\affiliation{Birla Institute of Technology and Science, Pilani, K K Birla Goa Campus, Goa 403726, India}

\begin{abstract}
Finite matrix-product-state sweep algorithms, such as the density-matrix renormalization group (DMRG) and the
time-dependent variational principle (TDVP), store one environment tensor per site. At the bond dimensions
needed for frustrated and two-dimensional systems this stack, rather than the state itself, exhausts the memory
of the device. We adapt checkpointing to these sweeps: every $k$-th environment is kept, with $k\approx\sqrt{N}$
for $N$ sites, and the blocks in between are rebuilt when the sweep returns to them. The rebuild repeats the
original contractions, so energies and states are reproduced bit for bit when the contractions are deterministic.
Environment memory falls by 4.6--10$\times$ on cylinders of 96--384 sites. Bond-dimension searches with an
implementation that rebuilds checkpoints at each half-sweep give gains of 1.77 for one level and 2.07 for two
levels with eight contraction chunks on a single 141\,GB GPU, using a 144-site triangular $J_1$-$J_2$ cylinder
and stored baselines with matching chunking. Separate timings with checkpoints retained across half-sweeps give
a one-level overhead of 6--15\%, depending on the local eigensolver. At $D=1536$ the two-level schedule with
eight chunks takes 1.42 times the sweep time of the equally chunked stored baseline, or 1.79 times that of the
unchunked stored baseline. Simple models of the memory and the run time reproduce the measurements to within a
few per cent. Unlike offloading, the method needs no host memory or disk,
and the two can be combined. Implementations are provided for ITensorMPS.jl, TeNPy, MPSKit.jl and quimb.
\end{abstract}

\maketitle

\section{Introduction}
\label{sec:intro}

The density-matrix renormalization group (DMRG)~\cite{White1992,White1993,Schollwock2005,Schollwock2011} and its
time-dependent extensions are central tools for strongly correlated quantum systems. Matrix-product-state (MPS)
methods for time evolution include time-evolving block decimation and adaptive time-dependent
DMRG~\cite{Vidal2004,WhiteFeiguin2004,Daley2004}, as well as the time-dependent variational principle
(TDVP)~\cite{Haegeman2011,Haegeman2016,Paeckel2019}. These methods address one-dimensional systems,
two-dimensional cylinders~\cite{Stoudenmire2012} and strongly correlated molecules~\cite{ChanSharma2011}.
They also provide classical benchmarks for quantum algorithms and starting states for quantum state
preparation and phase estimation~\cite{Ran2020,Lee2023}.

The bond dimension $D$ sets how much entanglement an MPS can carry across a cut: at most $D$ Schmidt
coefficients are kept, so the entanglement entropy is bounded by $\log D$. Ground states of gapped local
Hamiltonians obey an area law~\cite{VerstraeteCirac2006,Eisert2010}, which in one dimension means a bounded
entropy and explains the success of DMRG there. On a cylinder the cut runs around the circumference, the entropy
grows in proportion to the width, and the bond dimension needed for a given accuracy grows roughly exponentially
with it~\cite{Stoudenmire2012}. The open questions of frustrated magnetism, such as whether a candidate spin
liquid survives as the cylinder is widened, are therefore decided at bond dimensions in the thousands, and every
factor gained in $D$ is spent on the next width or on a tighter extrapolation. Out of equilibrium the demand is
steeper still, because the entanglement generated by a quench keeps growing with time~\cite{Paeckel2019}.

At these bond dimensions the dominant memory cost is not the MPS but the environment tensors. A left or right
environment is the contraction of the Hamiltonian with the part of the state that lies outside the sites being
updated. For a matrix product operator (MPO) of bond dimension $W$, a dense environment holds $D^2W$ numbers,
against $D^2d$ for an MPS tensor with local dimension $d$. A two-dimensional Hamiltonian wrapped onto a chain has
$W$ of several tens, so the stack of $N$ environments outweighs the state by an order of magnitude. For a 144-site
triangular $J_1$-$J_2$ cylinder with $W=38$ at $D=2048$ the stack occupies about 158\,GB in double precision,
more than a single NVIDIA H200 provides. The established remedies move environments to host memory or disk, or
distribute the calculation over several devices~\cite{offload1,offload2,offload3,offload4,ZhaiChan2021}.
Recomputing environments on demand has been noted as a further option when device and host memory are both
exhausted~\cite{recompute_future}.

Trading storage for recomputation is standard practice in reverse-mode automatic differentiation. Checkpointing
schedules~\cite{Griewank1992,GriewankWalther2000} reduce the memory of a backward pass from linear to sublinear
in the number of steps; they are used in training large neural networks~\cite{Chen2016} and in differentiable
tensor-network programming~\cite{Liao2019}. A DMRG or TDVP sweep has the same structure: one half-sweep produces
a sequence of environments and the next consumes them in reverse order. Here we carry the idea over to these
sweeps. The schedules themselves are standard. What is specific to a sweep is that checkpoints released on one
side are replaced by new ones on the other, and that the cost of a rebuild is weighed against a local eigensolver
instead of a gradient step.

We show that keeping a sparse set of environments and rebuilding the rest leaves the local update, and hence the
result, unchanged. Earlier searches with checkpoints rebuilt at each half-sweep increased the largest bond
dimension that fits on one GPU by factors of 1.77 for one level and 2.07 for two levels, relative to stored
baselines with matching chunking. We distinguish these limits from the fixed-$D$ measurements with retained
checkpoints in Sec.~\ref{sec:reach}. We give models for the memory and
the run time of the one-level schedule; they agree with measurements to within 2\% near the optimal spacing and
to within 1\%, respectively. We measure both quantities on GPUs with 24 to 141\,GB of usable memory and compare
recomputation with offloading to host memory and disk. Implementations in four libraries show how the same
schedule behaves with block-sparse tensors and under different memory managers.

\section{Environment checkpointing in MPS sweeps}
\label{sec:method}

An environment summarises the rest of the lattice as it appears to the sites being updated; it descends from the
environment block of the original formulation of DMRG~\cite{White1992,White1993}. It is a function of the MPS and
MPO tensors on one side of the active sites and of nothing else. If those tensors have not changed, contracting
them again returns the same environment, so nothing about the quantum state is lost by discarding it. A sweep
guarantees exactly this condition: the tensors ahead of the active sites are not modified until the sweep reaches
them.

\subsection{Reconstructing environments during a sweep}

Consider a finite MPS on $N$ sites with local Hilbert-space dimension $d$, and a Hamiltonian MPO of bond
dimension $W$. The MPS bond dimension $D$ limits the variational state, whereas $W$ describes the operator
representation~\cite{CrosswhiteBacon2008,Hubig2017}. A sweep consists of two half-sweeps, from left to right and
back. In a two-site DMRG step at $(j,j+1)$, the effective Hamiltonian is obtained from the left environment
$L_{j-1}$, the two local MPO tensors $W_j,W_{j+1}$ and the right environment
$R_{j+2}$~\cite{White1993,Schollwock2011,McCulloch2007}. A dense environment, which we also call a block, has
three indices of dimensions $D\times W\times D$. Moving the update one site to the right extends the left
environment by one contraction, $L_j=\mathcal{E}_j(L_{j-1})$, at cost
$c_\mathrm{env}=O(D^3Wd+D^2W^2d^2)$. Here $\mathcal{E}_j$ contracts the previous environment with the local
MPS tensor, its conjugate and the MPO tensor. The next half-sweep consumes these environments in reverse order,
which is why codes normally keep the whole stack.

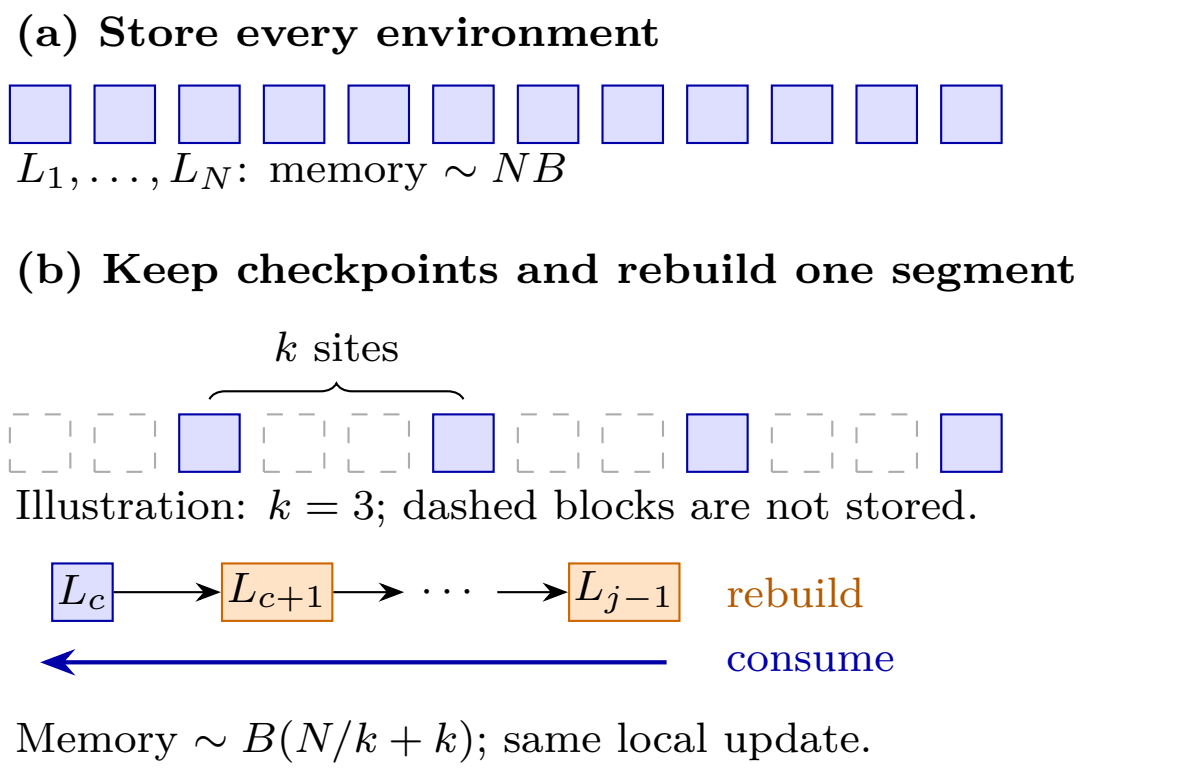
\begin{figure}[tbp]
\centering
\begin{tikzpicture}[x=0.56cm,y=0.58cm,
  stored/.style={draw=blue!65!black,fill=blue!13,minimum width=0.40cm,minimum height=0.38cm,inner sep=0pt},
  rebuilt/.style={draw=orange!80!black,fill=orange!22,minimum width=0.40cm,minimum height=0.38cm,inner sep=0pt},
  absent/.style={draw=gray!65,dashed,fill=white,minimum width=0.40cm,minimum height=0.38cm,inner sep=0pt},
  font=\footnotesize,>=Stealth]
\node[anchor=west,font=\footnotesize\bfseries] at (0,0.9) {(a) Store every environment};
\foreach \i in {1,...,12} {\node[stored] at (\i-0.5,0) {};}
\node[anchor=west] at (0,-0.65) {$L_1,\ldots,L_N$: memory $\sim NB$};
\node[anchor=west,font=\footnotesize\bfseries] at (0,-1.8) {(b) Keep checkpoints and rebuild one segment};
\foreach \i in {1,2,4,5,7,8,10,11} {\node[absent] at (\i-0.5,-3.75) {};}
\foreach \i in {3,6,9,12} {\node[stored] at (\i-0.5,-3.75) {};}
\draw[decorate,decoration={brace,amplitude=3pt}] (2.5,-3.25)--(5.5,-3.25)
 node[midway,above=4pt] {$k$ sites};
\node[anchor=west] at (0,-4.5) {Illustration: $k=3$; dashed blocks are not stored.};
\node[stored] (c) at (1,-5.45) {$L_c$};
\node[rebuilt,minimum width=0.73cm] (c1) at (3.3,-5.45) {$L_{c+1}$};
\node at (5.35,-5.45) {$\cdots$};
\node[rebuilt,minimum width=0.73cm] (jm) at (7.4,-5.45) {$L_{j-1}$};
\draw[->] (c)--(c1);
\draw[->] (c1)--(4.8,-5.45);
\draw[->] (5.9,-5.45)--(jm);
\node[anchor=west,text=orange!70!black] at (8.4,-5.45) {rebuild};
\draw[->,blue!65!black,thick] (7.9,-6.25)--(0.5,-6.25);
\node[anchor=west,text=blue!65!black] at (8.4,-6.25) {consume};
\node[anchor=west] at (0,-7.15) {Memory $\sim B(N/k+k)$; same local update.};
\end{tikzpicture}
\caption{Environment storage and reconstruction. (a) A full stack retains a block at every position.
(b) Checkpoints are spaced by $k$ sites. On a right-to-left half-sweep, the temporary blocks between a retained
$L_c$ and the required $L_{j-1}$ are rebuilt from left to right and then consumed in the opposite order.
Blue boxes denote checkpoints and orange boxes the active rebuilt segment. Used left blocks are released
while right checkpoints are produced for the next half-sweep. Boxes represent environment blocks, not physical
sites or MPS tensors.}
\label{fig:scheme}
\end{figure}

Instead, we keep every $k$-th environment as a checkpoint. During the return half-sweep, a missing block is
rebuilt from the nearest checkpoint on the unvisited side, and the segment in between is retained until its blocks
have been consumed (Fig.~\ref{fig:scheme}). The tensors on the unvisited side have not yet been updated in that
half-sweep, so the rebuild uses the same inputs as the original contraction. Consumed checkpoints are released,
and the checkpoints constructed on the opposite side are retained for the following half-sweep.

The checkpoints are a by-product of the ordinary sweep. A left-to-right half-sweep builds the left environments
and the return half-sweep builds the right ones, and retaining every $k$-th block adds no contraction to either.
The extra work is the rebuild of the dropped environments when they are needed in reverse order.
Algorithm~\ref{alg:sweep} gives a right-to-left half-sweep; the left-to-right version is its mirror image.

\begin{figure}[tbp]
\refstepcounter{algorithm}
\label{alg:sweep}
\footnotesize
\hrule\vspace{4pt}
\textbf{Algorithm~\thealgorithm.} Checkpointed right-to-left half-sweep
\par\vspace{4pt}\hrule\vspace{4pt}
\begin{tabular}{@{}r@{\hspace{0.5em}}p{0.91\columnwidth}@{}}
 & \emph{Input:} left checkpoints from the previous half-sweep, the left boundary $L_0$, and the right boundary $R_{N+1}$.\\
1 & \textbf{for} $j=N-1,\ldots,1$:\\
2 & \quad \textbf{if} $L_{j-1}$ is absent:\\
3 & \qquad choose the nearest stored $L_c$ with $c<j-1$;\\
4 & \qquad build and store $L_i=\mathcal{E}_i(L_{i-1})$ for $i=c+1,\ldots,j-1$.\\
5 & \quad solve with $L_{j-1},W_j,W_{j+1},R_{j+2}$ and split the updated two-site tensor;\\
6 & \quad build $R_{j+1}$ from $R_{j+2}$ and the updated tensor at site $j+1$;\\
7 & \quad release the consumed $L_{j-1}$, except $L_0$;\\
8 & \quad release $R_{j+2}$ unless it is a checkpoint or a boundary block.\\
\end{tabular}
\par\vspace{4pt}\hrule
\end{figure}

Because the rebuild repeats the same floating-point contractions, energies and states are reproduced bit for bit
whenever the contractions, the local solver and the remaining numerical settings are deterministic. No additional
truncation of the MPS is introduced. The same holds for TDVP, including the backward evolution of the bond
tensor, which uses the same environments~\cite{Haegeman2016,Paeckel2019}; its time-step and projection errors
remain those of the underlying integrator.

\subsection{Memory and recomputation cost}
\label{sec:model}

Let $B=D^2W$ be the number of entries of one dense environment in the bulk. We express environment memory in
these units; conversion to bytes multiplies by the storage per number. During a right-to-left half-sweep the left
checkpoints ahead of the active sites are consumed one by one, while every $k$-th right environment built behind
them is kept for the return. The two sets together number about $N/k$ at every moment. To these is added one
active segment of at most $k$ rebuilt blocks, so that, apart from boundary blocks and those in current use,
\begin{equation}
  M_\mathrm{env}(k) \approx B\left(\frac{N}{k}+k\right).
  \label{eq:mem}
\end{equation}
This is minimal at $k=\sqrt{N}$, where it equals $2\sqrt{N}B$ instead of $NB$. Increasing $k$ leaves fewer
checkpoints but lengthens the temporary segment, and the optimum balances the two. The minimum is shallow: any $k$
between $\sqrt{N}/2$ and $2\sqrt{N}$ is within 25\% of it, so $k$ needs no tuning. Equation~\eqref{eq:mem} treats
all blocks as equal, whereas near open ends the bond dimension, and with it the block, is smaller than in the
bulk; we return to this point in Sec.~\ref{sec:memory}.

All but one environment in each segment is rebuilt once per half-sweep, about $N(1-1/k)$ contractions, so the
time for a full sweep is
\begin{equation}
  t(k) = t_0 + 2N\left(1-\frac{1}{k}\right)c_\mathrm{env},
  \label{eq:time}
\end{equation}
where $t_0$ is the time with all environments stored. For large $k$ this is one extra environment update per site
and half-sweep. How much that matters depends on the local eigensolver, which applies the effective Hamiltonian
repeatedly, by Davidson~\cite{Davidson1975} or Lanczos iteration. To estimate the additional arithmetic work,
we count multiplications in the dense contractions. One application to a two-site tensor requires
$C_\mathrm{mv}=2D^3Wd^2+2D^2W^2d^3$ multiplications and one environment update
$C_\mathrm{env}=2D^3Wd+D^2W^2d^2$. Their ratio is $d(1+\epsilon)$, where
$\epsilon=Wd/(2D+Wd)$ is about 4\% at $D=1024$, $W=38$ and $d=2$. For a step with
$n_\mathrm{mv}$ applications, let $\sigma_\mathrm{op}\ge0$ denote the multiplication count of the decomposition
and remaining work in units of $C_\mathrm{env}$. The regular environment update contributes one further unit.
Within this count, the relative increase in arithmetic work is
\begin{equation}
  \delta_\mathrm{op} \approx
  \frac{1-1/k}{d(1+\epsilon)n_\mathrm{mv}+1+\sigma_\mathrm{op}}
  \;\le\; \frac{1-1/k}{d\,n_\mathrm{mv}+1}.
  \label{eq:delta}
\end{equation}
For spin-$1/2$ sites and $k=12$, the right-hand side is about 10\% with four applications per solve and 18\%
with two. With fifty applications it is below 1\%. These are operation-count estimates, not rigorous bounds on
wall time. The contractions and the decomposition can achieve different throughputs, and data movement and
allocation also contribute to the measured time. If the dominant contractions have comparable throughput, the
count explains why a more expensive local solve reduces the relative runtime overhead. At fixed $W$, $d$ and
solver settings, their leading multiplication counts scale as $D^3$, so the estimate has little dependence on
$D$. In a single-site update the two contraction counts are equal and the factor $d$ is absent, which increases
the estimated relative arithmetic overhead. Runtime comparisons below use measured sweep times.

The reversal problem of automatic differentiation traverses a sequence of steps backwards with a fixed number of
snapshots, and binomial schedules, implemented by \texttt{revolve}, minimise the recomputation in that
setting~\cite{Griewank1992,GriewankWalther2000}. They gain by placing new snapshots in the slots that are freed as
the reversal proceeds. In a sweep with retained checkpoints (Algorithm~\ref{alg:sweep}) this freedom is absent. The
preceding half-sweep supplies the checkpoints at no cost, and every slot freed on one side is taken up at once by a
checkpoint laid down on the other, so the number of checkpoints stays at about $N/k$ throughout. Uniform spacing is
therefore the natural choice, and it needs only a few dozen lines in each library. Only if the checkpoints are
rebuilt at the start of every half-sweep, at the cost of a further $N$ environment updates, do the freed slots stay
free; gaps that widen as the checkpoints are consumed would then lower the memory from $2\sqrt{N}B$ to about
$\sqrt{2N}\,B$.

\subsection{Two-level schedule}
\label{sec:twolevel}

If one level does not save enough memory, a second can be added~\cite{GriewankWalther2000}. Every $k_1$-th
environment is kept globally, as before. When a segment of $k_1$ blocks is rebuilt, only every $k_2$-th block in
it is kept, and each of the remaining blocks is rebuilt from the nearest kept block when the sweep reaches it and
released after use. The active segment then holds $k_1/k_2$ blocks instead of $k_1$, so that
\begin{equation}
  M_\mathrm{env}(k_1,k_2)\approx B\left(\frac{N}{k_1}+\frac{k_1}{k_2}\right),
  \label{eq:mem2}
\end{equation}
plus the block in current use. For fixed $k_2$ this is minimal at $k_1=\sqrt{Nk_2}$, where it equals
$2\sqrt{N/k_2}\,B$: the second level acts as a one-level schedule on a chain coarse-grained by $k_2$. The price
is paid in recomputation. A block at distance $m$ from the nearest kept block needs $m$ updates, so a group of
$k_2$ blocks costs $1+2+\dots+(k_2-1)=k_2(k_2-1)/2$ updates, or $(k_2-1)/2$ per site, in addition to the one
update per site of the first pass through the segment. The sweep time becomes
\begin{equation}
  t(k_1,k_2) \approx t_0 + N\,(k_2+1)\,c_\mathrm{env}.
  \label{eq:time2}
\end{equation}
For $N=144$ the one-level optimum holds about 24 blocks. The choice $k_1=36$, $k_2=6$ used below holds about 10,
with 3.3 extra updates per site and half-sweep on an exact count, against 0.9 for one level. Keeping the whole
group of $k_2$ blocks instead would reduce this to about two updates per site at the price of $k_2$ further
blocks. We use the leaner variant because the second level is needed only when memory is the binding constraint.

\subsection{Where the saving applies}
\label{sec:when}

The peak memory is $M_\mathrm{other}+M_\mathrm{env}$, where $M_\mathrm{other}$ comprises the MPS, the
intermediates of the effective Hamiltonian and the basis vectors of the eigensolver. Checkpointing reduces only
the second term, so it helps most where that term dominates: long systems, wide MPOs and large bond dimensions.
Where working memory dominates, it has to be reduced by other means. A more compact
MPO~\cite{Hubig2017,Chan2016} or the use of symmetry sectors~\cite{McCulloch2007,Singh2010} acts on other terms
of this balance and combines with checkpointing. Since every term scales as $D^2$ in the bulk, the largest bond
dimension that fits a given device scales as the inverse square root of the memory needed at fixed $D$. A
reduction of the peak by a factor $r$ therefore raises the attainable bond dimension by about $\sqrt{r}$.

\section{Implementations and correctness checks}
\label{sec:impl}

The implementations cover two situations: a code with explicit control of device memory, and existing
tensor-network libraries with their own allocation behaviour. Our main measurements use a standalone CUDA code
for two-site DMRG (double precision, contractions with cuBLAS and cuTENSOR, no quantum-number symmetries). Two
changes besides checkpointing were needed before the environment saving became visible in the peak. The Davidson
basis is sized to the actual number of matrix-vector products, and the intermediate of the effective Hamiltonian
can be split into chunks that are contracted one after another, which we call chunking. Device memory is managed
with the stream-ordered allocator (\texttt{cudaMallocAsync}/\texttt{cudaFreeAsync}), so that released
environments return to the pool at once. Memory is quoted in decimal gigabytes ($10^9$ bytes). In these units the
NVIDIA H100 and H200 cards used here report 85.0 and 150.1\,GB; we refer to them by their nominal sizes of 80 and
141\,GB.

We also implemented the scheme as drop-in components for four libraries:
\begin{itemize}
\item ITensorMPS.jl~\cite{ITensor}: \texttt{CheckpointedProjMPO}, an \texttt{AbstractProjMPO} usable with
  \texttt{dmrg} and \texttt{tdvp}, with an optional two-level schedule and offload of checkpoints to host memory
  or disk;
\item TeNPy~\cite{TeNPy}: a checkpointed MPO environment, now part of the library as the DMRG option
  \texttt{env\_checkpoint\_every} and available as a plugin for earlier versions;
\item MPSKit.jl~\cite{MPSKit}: an environment type that uses MPSKit's own dependency tracking;
\item quimb~\cite{quimb}: a checkpointed moving environment for open boundary conditions.
\end{itemize}
None requires changes to user code beyond selecting the environment type. The code is released under the MIT
License~\cite{envcheckpoint}. The GPU code was built with CUDA~12.1 and cuTENSOR~2; the library benchmarks used
ITensorMPS.jl~0.4.2 with Julia~1.10, MPSKit.jl~0.13.13, TeNPy~1.1.1 with NumPy~2.4.6 (OpenBLAS~0.3.31) and
quimb~1.15.0.

For the GPU benchmarks we use the spin-$1/2$ $J_1$-$J_2$ Heisenberg Hamiltonian,
\[
H=J_1\sum_{\langle i,j\rangle}\mathbf{S}_i\cdot\mathbf{S}_j
 +J_2\sum_{\langle\!\langle i,j\rangle\!\rangle}\mathbf{S}_i\cdot\mathbf{S}_j,
\]
where $\mathbf{S}_i$ is the spin operator at site $i$ and the sums run over nearest and next-nearest neighbours;
we set $J_1=1$. On the square lattice we take $J_2/J_1=0.5$, in the non-magnetic window between N\'eel and stripe
order~\cite{Gong2014}, and on the triangular lattice $J_2/J_1=0.12$, inside the proposed spin-liquid
phase~\cite{ZhuWhite2015}. In both regimes competing states lie close in energy, and the conclusions drawn from
cylinders rest on the largest bond dimensions available, which makes these models natural test cases. Cylinders
are open along their length and periodic around the circumference. They are labelled by the number of sites
around the circumference and by the orientation of the lattice, with a nearest-neighbour bond direction along the
circumference (YC) or along the axis (XC). We use square YC6, YC8 and YC10 cylinders ($W=20$, 26 and 32) and
triangular XC6 and XC8 cylinders ($W=38$ and 50), with up to 384 sites. The spin-1 and quantum-chemistry benchmarks are specified where
they appear.

The numerical checks run the same initial state, local solver and truncation settings with the two storage
schemes. Checkpointing alters neither the variational optimisation nor the basis enrichment used by a
library~\cite{White2005,Hubig2015}, so agreement between the two runs tests the reconstruction itself, whether or
not the chosen bond dimension has converged the physics.

In the GPU code, energies are identical to the last bit between stored and checkpointed runs for every $k$ from 4
to 72, for the two-level schedule, for three models and two eigensolvers. Over a 20-sweep run at $D=1536$ the
maximum difference is exactly zero. Chunking the effective Hamiltonian changes the order of summation and hence
the last two to three digits, an agreement to $\sim\!10^{-9}$. In eight runs at $D=1024$ that varied $k$, the
Davidson basis size and the chunking, checkpointing changed no digit of the energy, the truncation error or the
residual.

In the libraries, DMRG energies agree exactly (difference $0.0$) with those of the native environments in
ITensorMPS.jl (dense and U(1)), TeNPy (including mixers and disk caching), MPSKit.jl (dense and U(1)) and quimb,
and TDVP states agree to $1-|\langle a|b\rangle|\sim 10^{-15}$. In ITensorMPS.jl, one-site TDVP agrees to
$1-|\langle a|b\rangle|=7\times10^{-16}$ and the two-level schedule ($k_1=7$, $k_2=3$) reproduces the energy
exactly. For infinite MPS in TeNPy the option is inactive. In the longer CPU benchmarks of Sec.~\ref{sec:cpu}
(real-time TDVP over 13 steps, ab initio DMRG with $W=578$, MPSKit.jl DMRG up to $D=768$) all printed energies
agree to 12 digits.

\section{Memory savings and computational performance}
\label{sec:performance}

We separate three questions: what recomputation costs in time at fixed numerical settings, how much memory it
saves, and how much further in bond dimension a given device can then go.

\subsection{Runtime overhead}
\label{sec:cost}

Two local solvers are used throughout. Davidson-4 limits the Davidson iteration to four matrix-vector products
per local solve and truncates by diagonalising the reduced density matrix $\rho$ of the two-site
tensor~\cite{White1992}. Davidson-2 uses two products and a warm-started truncation: the kept basis $X$ is not
obtained from a diagonalisation but is initialised with the MPS tensor that is being replaced and improved by one
step of subspace iteration, $X\leftarrow\mathrm{qr}(\rho X)$. The full diagonalisation is used only where the bond
dimension changes.
Local solves are thus not converged within a step, as is common in production runs, where the sweeping itself
converges the state. Stored and checkpointed runs always share the same setting.

Table~\ref{tab:cost} lists the overhead of the one-level schedule with $k=12$. The cleanest comparison is with a
stored run that keeps its environments from one half-sweep to the next, the checkpointed run keeping its
checkpoints in the same way. For the triangular XC6 cylinder at $D=1024$ the overhead is then 6.0\% with
Davidson-4 and 13.7\% with Davidson-2, with a spread of at most 0.2\% over three runs. The larger overhead
for the cheaper solver follows the trend of the operation-count estimate in Eq.~\eqref{eq:delta}. Energies agree to the last printed digit in all twelve runs, and the peak memory
is 19.3 against 52.1\,GB. At $D=1536$ the Davidson-2 overhead is 15\% (Sec.~\ref{sec:reach}), consistent with
an overhead that does not depend on $D$. With local solves converged fully it falls to 0.5\% (693 against 690\,s
per sweep).

An earlier set of runs used a stored baseline that rebuilds all environments at the start of each half-sweep,
with the checkpoints rebuilt likewise. Both variants then carry the same additional rebuild, and the extra work
of checkpointing is unchanged at $N(1-1/k)$ environment updates per half-sweep. These runs give median overheads
of 6.2\% (triangular XC6, $D=1024$) and 6.4\% (square YC8, $D=2048$) with Davidson-4, and 14.5\% with Davidson-2,
with a larger scatter between runs (Table~\ref{tab:cost}). Equation~\eqref{eq:time}, with $t_0$ and
$c_\mathrm{env}$ measured once, reproduces every measured $t(k)$ to within 1\%.

The extra work cannot be overlapped with the solver. Profiling shows the GPU 98--99\% busy during a sweep, so the
rebuild competes for a device that is already saturated.

\begin{table}[tbp]
\caption{Overhead in time of checkpointing ($k=12$) relative to storing all environments, GPU code on one H200;
energies are identical in every pair. Davidson-4: at most four matrix-vector products per local solve and
truncation through the reduced density matrix. Davidson-2: two products and warm-started truncation. In the first
two rows, stored runs keep their environments and checkpointed runs their checkpoints across half-sweeps; in the
last three, both are rebuilt at the start of each half-sweep. IQR: interquartile range.}
\label{tab:cost}
\begin{ruledtabular}
\begin{tabular}{@{}lccc@{}}
model, $D$ & solver & median & IQR (runs)\\
\hline
tri. XC6, 1024 & Davidson-4 & 6.0\% & 5.98--6.00\% (3)\\
tri. XC6, 1024 & Davidson-2 & 13.7\% & 13.5--13.7\% (3)\\
\hline
tri. XC6, 1024 & Davidson-4 & 6.2\% & 4.8--8.4\% (10)\\
square YC8, 2048 & Davidson-4 & 6.4\% & 6.2--6.5\% (5)\\
tri. XC6, 1024 & Davidson-2 & 14.5\% & 12.1--21.6\% (10)\\
\end{tabular}
\end{ruledtabular}
\end{table}

\subsection{Memory}
\label{sec:memory}

\begin{figure}[tbp]
\includegraphics[width=\columnwidth]{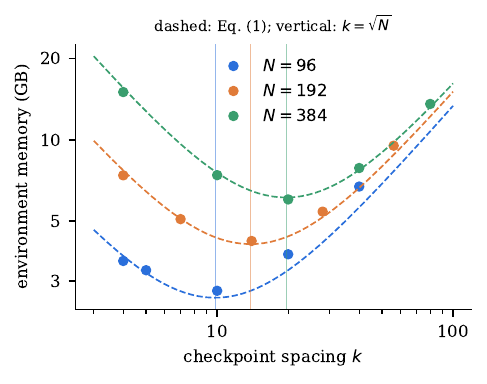}
\caption{Environment memory against checkpoint spacing $k$ for square YC6 cylinders at $D=1024$ (points), with
Eq.~\eqref{eq:mem} for $B$ taken from the run that stores all environments (dashed). Vertical lines mark
$k=\sqrt{N}$.}
\label{fig:k}
\end{figure}

Table~\ref{tab:length} shows peak and environment memory for square YC6 cylinders of increasing length at
$D=1024$. The spacing that minimises memory in a scan over $k$ is 10, 14 and 20 for $N=96$, 192 and 384, against
$\sqrt{N}=9.8$, 13.9 and 19.6. Environment memory falls by factors of 4.6, 6.7 and 10, close to the $\sqrt{N}/2$
of Eq.~\eqref{eq:mem}, and the saving in peak memory grows from 1.9 to 3.8. Figure~\ref{fig:k} shows the full
scans. With $B$ and the non-environment memory taken from the stored run, Eq.~\eqref{eq:mem} predicts the peak of
all 15 checkpointed runs to within 2\% for $k\le2\sqrt{N}$. For larger $k$ it underestimates the peak by up to
7\% ($k=40$, $N=96$). The block size inferred from the stored run is an average that includes the small blocks
near the open ends, whereas a long active segment in the bulk consists of full-size blocks.

\begin{table}[tbp]
\caption{Square YC6 cylinder, $D=1024$, GPU code: memory with all environments stored and with checkpointing at
the measured optimal $k$.}
\label{tab:length}
\begin{ruledtabular}
\begin{tabular}{@{}rcccc@{}}
$N$ & best $k$ ($\sqrt{N}$) & peak (GB) & env. (GB) & saving\\
\hline
96 & 10 (9.8) & 21.3 $\to$ 11.3 & 12.7 $\to$ 2.8 & 1.9$\times$\\
192 & 14 (13.9) & 38.6 $\to$ 14.4 & 28.4 $\to$ 4.2 & 2.7$\times$\\
384 & 20 (19.6) & 73.2 $\to$ 19.4 & 59.8 $\to$ 6.0 & 3.8$\times$\\
\end{tabular}
\end{ruledtabular}
\end{table}

The saving also depends on the model, through the length $N$ and the MPO bond dimension $W$, whose product sets
the size of the stack. At $D=1024$ the peak falls by a factor of 1.5 for square YC10 ($N=80$, $W=32$), 1.8 for
square YC8 ($N=128$, $W=26$), 2.7 for triangular XC6 ($N=144$, $W=38$) and 3.4 for triangular XC8 ($N=192$,
$W=50$; 86.1\,GB $\to$ 25.1\,GB). The ordering follows $NW$. The YC10 cylinder has a wider MPO than YC8 but is
shorter, and gains less.

\begin{figure}[tbp]
\includegraphics[width=\columnwidth]{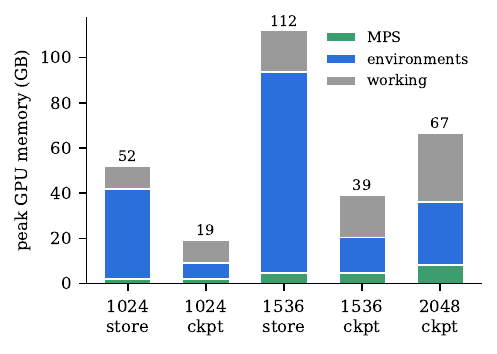}
\caption{Peak GPU memory split into MPS, environments and working memory for the triangular XC6 cylinder, with
all environments stored (store) and checkpointed with $k=12$ (ckpt). Storing all environments at $D=2048$ does
not fit on the device.}
\label{fig:breakdown}
\end{figure}

Two settings unrelated to checkpointing also change the peak, and we varied them separately (triangular XC6,
$D=1024$, Davidson-4). With a Davidson basis of eight vectors and no chunking, checkpointing with $k=12$ lowers
the peak from 52.0 to 19.2\,GB and raises the sweep time from 32.3 to 34.0\,s. A basis of 160 vectors adds
10\,GB, and four chunks save 1.9\,GB for 14\% more time, with or without checkpointing. Chunking becomes cheaper
relative to the sweep as $D$ grows. At $D=2048$ (Davidson-2, $k=12$), two, four, eight and sixteen chunks save
5.1, 7.5, 8.9 and 9.4\,GB of a 66.0\,GB peak for 1\%, 3\%, 13\% and 25\% more time, so four chunks capture most
of the saving at little cost. Figure~\ref{fig:breakdown} shows how the peak divides between MPS, environments and
working memory.

\subsection{Largest attainable bond dimension}
\label{sec:reach}

The practical question is how far a given device can go. We define the limit as the largest $D$ for which two
full sweeps complete, located on a grid of 64 or 128. Table~\ref{tab:config} separates two sets of measurements
for the triangular XC6 cylinder on one H200. The bond-dimension searches use a build that rebuilds
environments or checkpoints at the start of each half-sweep. The fixed-$D$ timings and memory measurements use
a build with environments or checkpoints retained across half-sweeps.

In the earlier searches, without chunking, the one-level schedule raised the limit from 1664 to 2944, a factor
of 1.77. With eight chunks the stored limit was 1792 and the two-level schedule ($k_1=36$, $k_2=6$) reached 3712,
a factor of 2.07. On a grid of 128 these ratios are defined to about $\pm0.1$. For the retaining build, the
measured peaks at $D=1536$ give the square-root estimates $\sqrt{111.9/39.1}=1.69$ and
$\sqrt{106.9/25.4}=2.05$ from Sec.~\ref{sec:when}, close to the ratios measured with the rebuilding build.

\begin{table}[tbp]
\caption{Triangular XC6 cylinder ($N=144$) on one H200, Davidson-2. The rebuilding-build columns report earlier
searches for the largest $D$ completing two full sweeps, on a grid of 128. Their ratios compare runs with the same
chunking. The retaining-build columns report time per sweep $t$ and peak memory $M$ at $D=1536$, averaged over
the second and third sweeps with environments or checkpoints kept across half-sweeps.}
\label{tab:config}
\begin{ruledtabular}
\begin{tabular}{@{}lcccc@{}}
 & \multicolumn{2}{c}{Rebuilding build} & \multicolumn{2}{c}{Retaining build}\\
configuration & $D_\mathrm{max}$ & ratio & $t$ (s) & $M$ (GB)\\
\hline
stored & 1664 & -- & 42.6 & 111.9\\
stored, 8 chunks & 1792 & -- & 53.7 & 106.9\\
one-level, $k=12$ & 2944 & 1.77 & 49.0 & 39.1\\
two-level, 8 chunks & 3712 & 2.07 & 76.3 & 25.4\\
\end{tabular}
\end{ruledtabular}
\end{table}

The fixed-$D$ timings distinguish the two runtime baselines. At $D=1536$ the retaining build gives
$49.0/42.6\approx1.15$ times the stored sweep time for one level without chunking. Eight chunks alone raise the
stored time by 26\%. The two-level schedule with eight chunks takes $76.3/53.7\approx1.42$ times the time of
the equally chunked stored run, or $76.3/42.6\approx1.79$ times that of the unchunked stored run. The 2.07
reach ratio is measured with the rebuilding build and the 1.42 runtime ratio with the retaining build, each
relative to an equally chunked stored baseline.
The count of Sec.~\ref{sec:twolevel} accounts for the additional time at matched chunking. The
one-level run gives $c_\mathrm{env}\approx24$\,ms, and the two-level schedule performs about 475 environment
updates per half-sweep against 132. This predicts 23\,s per sweep on top of the stored run with the same
chunking, and the measured difference is 22.6\,s. In memory,
the two-level run holds 7.2\,GB of environments, ten blocks of 0.72\,GB as expected from Eq.~\eqref{eq:mem2},
against 15.8\,GB (22 blocks) for one level and 88.8\,GB with all environments stored.

At the two-level limit measured with the earlier rebuilding build, the environments are no longer the largest term. Of the 138\,GB in use at $D=3712$, the
MPS takes 27\,GB, the environments 38\,GB and the working memory 74\,GB. Further gains must come from the working
memory of the local update.

Table~\ref{tab:reach} lists the earlier rebuilding-build searches on the other devices and memory budgets. Here the stored runs use no chunking and the
checkpointed runs the two-level schedule with eight chunks (triangular; $k_1=36$, $k_2=6$) or sixteen (square;
$k_1=32$, $k_2=6$), so the ratios include the effect of chunking. On the H200 that effect is one grid step in the
stored limit: the ratio is 2.23 with the chunking unmatched and 2.07 with it matched. On an 80\,GB H100 the
triangular limit rises from 1280 to 2816. For the square YC8 cylinder the search was stopped at 3328 on the H100,
where the run filled the card, and at 4096 on the H200 with 124.6\,GB in use, so those values are lower bounds.

To emulate cards with less memory we reserve a fixed ballast allocation on the H200 before the run and repeat the
search for the triangular cylinder. With 24\,GB usable the limit rises from 640 to 1408, with 40\,GB from 768 to
1920, and with 48\,GB from 896 to 2048. These are ratios of 2.2 to 2.5, with an uncertainty of about $\pm0.3$
from the grid of 128. At the 24\,GB limit the environments occupy 15.7\,GB stored at $D=640$ but 5.4\,GB
checkpointed at $D=1408$. At $D=768$ a checkpointed sweep takes 14.1\,s against 7.6\,s stored. The stored runs
start from a random state and the checkpointed runs from a converged $D=1280$ state; a checkpointed run from a
random state completes at 24\,GB and $D=1408$ with the same 5.4\,GB of environments.

\begin{table}[tbp]
\caption{Largest bond dimension completing two full sweeps on one GPU, GPU code. Stored runs use no chunking;
checkpointed runs use the two-level schedule with chunking (see text). These earlier rebuilding-build limits
are located on a grid of 64 or 128.
In the last three rows the usable memory of an H200 is capped by a ballast allocation. Table~\ref{tab:config}
resolves the H200 triangular case by configuration.}
\label{tab:reach}
\begin{ruledtabular}
\begin{tabular}{llccc}
GPU & model & stored & ckpt. & ratio\\
\hline
H100 80\,GB & tri. XC6 & 1280 & 2816 & 2.2\\
H100 80\,GB & sq. YC8 & 1664 & $\ge 3328$ & $\ge 2.0$\\
H200 141\,GB & tri. XC6 & 1664 & 3712 & 2.2\\
H200 141\,GB & sq. YC8 & 2176 & $\ge 4096$ & $\ge 1.9$\\
\hline
H200, 24\,GB cap & tri. XC6 & 640 & 1408 & 2.2\\
H200, 40\,GB cap & tri. XC6 & 768 & 1920 & 2.5\\
H200, 48\,GB cap & tri. XC6 & 896 & 2048 & 2.3\\
\end{tabular}
\end{ruledtabular}
\end{table}

In ITensorMPS.jl with U(1) symmetry on an 80\,GB H100 (square YC8), the native \texttt{ProjMPO} reaches $D=3584$
and \texttt{CheckpointedProjMPO} reaches $D=4608$ on a coarse grid, a factor of 1.29, with comparable sweep
times. With dense tensors checkpointing does not extend the reach: every variant, including two-level schedules,
host offload of checkpoints, $k=2$ and chunked effective Hamiltonians, fails at $D=2048$ or 2560. The allocation
that fails is a single environment-sized temporary inside the two-site update. The limit is thus set by the
working memory of the library and not by the stored environments, the situation described in
Sec.~\ref{sec:when}. \texttt{DiskProjMPO} reaches 2560 but is slower: 614\,s per sweep at $D=1536$, against 357\,s
checkpointed and 376\,s for \texttt{ProjMPO}.

\subsection{Comparison with host-memory and disk offloading}
\label{sec:offload}

Moving environments off the device exchanges GPU memory for host storage and data transfer, and asynchronous
transfer and distributed storage are central to several large-scale
implementations~\cite{offload1,offload3,offload4,ZhaiChan2021}. We compared recomputation with asynchronous
offload and prefetching in the same code, on the triangular XC6 cylinder.

We first routed the offload through a file system (Table~\ref{tab:offload}). At $D=1024$ recomputation takes
18.2\,s per sweep. Offload to host memory through a RAM disk takes 25\,s. A node-local disk matches this only
while the page cache of the operating system absorbs the writes (25\,s in the first sweep, 208\,s afterwards).
The parallel file system takes 34--37\,s and synchronous offload 75--85\,s. At $D=2048$ the RAM-disk run
exceeded the 200\,GB memory limit of its job in the first sweep, while recomputation ran at 128\,s per sweep.

\begin{table}[tbp]
\caption{Time per sweep (s) for the triangular XC6 cylinder: recomputation ($k=12$) against asynchronous offload
with prefetching through a file system, GPU code on one H200, Davidson-2, environments rebuilt at the start of
each half-sweep.}
\label{tab:offload}
\begin{ruledtabular}
\begin{tabular}{lcc}
method & $D=1024$ & $D=2048$\\
\hline
recomputation & 18.2 & 128\\
host memory (\texttt{/dev/shm}) & 24.6--25.1 & out of memory\\
node-local disk & 24.9, then 208 & 130, then 1291\\
parallel file system & 34--37 & 188--221\\
same, synchronous & 75--85 & --\\
all environments stored & 16.2--16.3 & --\\
\end{tabular}
\end{ruledtabular}
\end{table}

A direct implementation does better (Table~\ref{tab:pinned}). Environments are copied with
\texttt{cudaMemcpyAsync} to and from page-locked (pinned) host buffers on a separate stream, with two blocks
fetched ahead of the sweep. After a first sweep of 25\,s that allocates the buffers, a sweep at $D=1024$ takes
16.3--16.4\,s. This equals the time with every environment stored and is 10\% less than with recomputation, at a
peak GPU memory of 13.6\,GB against 19.1\,GB checkpointed. At $D=1536$ the sweep time varies between 49 and
62\,s. At $D=2048$ the pinned run was terminated in its second sweep on exceeding the 290\,GB memory limit of the job,
while the checkpointed run completed at a peak of 66.6\,GB and 110--128\,s per sweep. Energies are identical in
every run.

The host requirement of both offload routes, as implemented here, is twice the stack. Each keeps one host
buffer, or one file, for every left and every right block and reuses it in the next sweep without releasing it in
between, so the left and the right stack reside on the host together. At $D=2048$ one stack is 158\,GB and the
two amount to about 315\,GB. An implementation that released the buffers of consumed blocks would need half of
this and would have fitted within the 290\,GB limit.

\begin{table}[tbp]
\caption{Pinned asynchronous offload compared with stored and checkpointed ($k=12$) environments: time per sweep
$t$ and peak GPU memory $M$ for the triangular XC6 cylinder on one H200, Davidson-2, three runs per setting from
the same initial state, other settings as in Table~\ref{tab:offload}.}
\label{tab:pinned}
\begin{ruledtabular}
\begin{tabular}{@{}lcccc@{}}
 & \multicolumn{2}{c}{$D=1024$} & \multicolumn{2}{c}{$D=1536$}\\
 & $t$ (s) & $M$ (GB) & $t$ (s) & $M$ (GB)\\
\hline
stored & 16.2--16.3 & 51.8 & 49.2 & 111.9\\
checkpointed & 18.1 & 19.1 & 55.4 & 39.0\\
pinned offload & 16.3--16.4 & 13.6 & 49--62 & 26.4\\
\end{tabular}
\end{ruledtabular}
\end{table}

The two methods thus draw on different resources. One block at $D=1024$, $W=38$ is 0.32\,GB; moving it over
PCIe~5 at about 25\,GB/s takes 13\,ms, against 8.3\,ms to rebuild it. The transfer, however, runs on the copy
engines while the solver computes, so a well-overlapped offload costs almost no time, whereas a rebuild competes
for the saturated GPU. The scaling favours offload further as $D$ grows, since the transfer time grows as $D^2$
and the rebuild as $D^3$. What offload requires instead is host memory for the whole stack, of order $NWD^2$,
which becomes the binding limit at large $D$ and on shared nodes with per-job memory limits. Recomputation needs
no host memory, and its footprint on the device grows only as $\sqrt{N}$. Where host memory is plentiful, pinned
offload is the faster choice; where it is not, checkpointing remains available. The two combine naturally: if
only the $N/k$ checkpoints leave the device, the host requirement falls by a factor of $k$ and the device holds a
single segment. The ITensorMPS.jl implementation offers this combination.

\subsection{CPU libraries}
\label{sec:cpu}

CPU implementations test how the schedule interacts with block-sparse tensors and with the memory management of
each library.

In TeNPy, on square YC6 cylinders with $S^z$ conservation, the peak resident memory falls by 2.9$\times$
($24\times6$, $D=512$), 3.3$\times$ ($24\times6$, $D=1024$), 4.0$\times$ ($48\times6$, $D=1024$) and 4.3$\times$
($48\times6$, $D=2048$; 49.7\,GB $\to$ 11.6\,GB), with identical energies. No run took longer with checkpointing,
and in three repeated pairs at $D=1024$ the checkpointed runs were 6--15\% faster, an effect examined in
Sec.~\ref{sec:faster}.

Time evolution is a natural application. After a quench the entanglement entropy typically grows linearly in
time, so the bond dimension needed for a faithful state grows exponentially~\cite{Vidal2004,WhiteFeiguin2004,Paeckel2019}
and a simulation ends when the state no longer fits. A fixed factor gained in $D$ then buys a fixed increment in
the reachable time. Checkpointing moves this storage limit and leaves the growth of entanglement and the errors
of the integrator untouched.

\begin{figure}[tbp]
\includegraphics[width=\columnwidth]{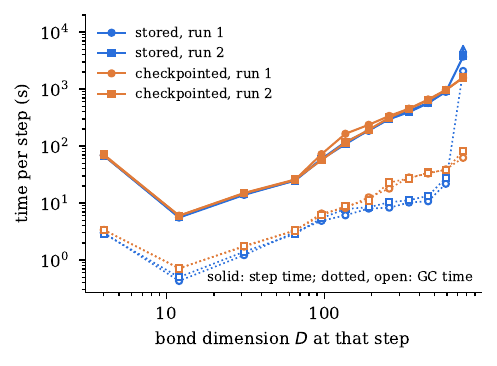}
\caption{Real-time two-site TDVP in ITensorMPS.jl (square Heisenberg YC6, $12\times6$, U(1), $\delta t=0.05$):
time per step (solid) and garbage-collection time per step (dotted) against the bond dimension reached at that
step, two runs each on a dedicated 64-core node. The first step includes compilation. The stored run 2 was stopped
by its time limit during the last step; the arrow marks the lower bound. Energies agree to all printed digits at
every step.}
\label{fig:tdvp}
\end{figure}

Table~\ref{tab:cpu} lists three further comparisons, each with one storage scheme per process. The first is
real-time two-site TDVP in ITensorMPS.jl: a quench from a N\'eel state on a square Heisenberg YC6 cylinder
($12\times6$, U(1), $\delta t=0.05$, $D\le768$). Energies agree at every step and the peak memory falls from 6.38
to 4.82\,GB. Figure~\ref{fig:tdvp} shows two runs of each scheme on a dedicated node (64-core AMD EPYC 7542, one
thread per Julia process, eight BLAS threads). Up to $D=589$ the checkpointed steps are slower, by 10\% summed
over steps 2--11, as expected from the extra environment updates. At $D=758$ the order reverses: the stored step
takes 3716\,s in one run and did not finish within the remaining 3870\,s of the other, while the checkpointed step
takes 1581 and 1606\,s. In a first run with four threads the corresponding step, at $D=768$, took 3969\,s stored
against 1369--1440\,s checkpointed, so that within the same 2.5\,h the checkpointed run completed 16 steps and
the stored run 13.

The second comparison is DMRG in MPSKit.jl on a spin-1 Heisenberg YC6 cylinder ($24\times6$, U(1)). The
checkpointed run is 1.22$\times$ faster overall and uses 7.63 instead of 9.35\,GB at $D=768$; two repeated pairs
with $D\le512$ on the dedicated node give 1.14$\times$ at equal peak memory. The third is ab initio
DMRG~\cite{ChanSharma2011,Chan2016} of an H$_{16}$ chain (STO-6G) in ITensorMPS.jl. Here the MPO is wide
($W=578$) and the chain short, so the effective Hamiltonian dominates the memory and checkpointing changes
little: 6.70 $\to$ 6.23\,GB, with the checkpointed run 2--9\% slower.

\begin{table}[tbp]
\caption{CPU benchmarks, stored $\to$ checkpointed environments (energies identical). TDVP and H$_{16}$:
ITensorMPS.jl. Rows marked $\dagger$ are medians over repeated pairs on a dedicated node; the others are single
runs. GC: garbage collection. TeNPy: see Table~\ref{tab:tenpy}.}
\label{tab:cpu}
\begin{ruledtabular}
\begin{tabular}{@{}lcc@{}}
run & time (s) & RAM (GB)\\
\hline
TDVP step, $D=768$ & 3969 $\to$ 1369 & 6.38 $\to$ 4.82\\
TDVP step, $D=758$ $\dagger$ & $\ge$3716 $\to$ 1594 & 5.5 $\to$ 4.5\\
same, no forced GC $\dagger$ & 3414 $\to$ 1734 & 5.6 $\to$ 4.6\\
MPSKit, $D\le768$ & 3003 $\to$ 2455 & 9.35 $\to$ 7.63\\
MPSKit, $D\le512$ $\dagger$ & 1753 $\to$ 1532 & 6.4 $\to$ 6.5\\
H$_{16}$, $D\le512$ & 232 $\to$ 253 & 6.70 $\to$ 6.23\\
H$_{16}$, $D\le512$ $\dagger$ & 412 $\to$ 426 & 6.8 $\to$ 6.3\\
\end{tabular}
\end{ruledtabular}
\end{table}

Codes that already keep their environments on disk, such as block2~\cite{block2}, are in a different position:
there the same schedule would save disk space and input/output instead of memory. For an H$_{50}$ chain (STO-6G,
50 orbitals, $D=1000$, SU(2)) the environment files of block2 occupy about 40\,GB, and disk input/output accounts
for 16--29\% of the run.

\subsection{Memory management and CPU timing}
\label{sec:faster}

On the GPU, checkpointing always costs time (Sec.~\ref{sec:cost}). On CPUs, in libraries written in
garbage-collected or allocation-heavy languages, we repeatedly found the checkpointed run as fast as the stored
one or faster (TeNPy, MPSKit.jl, and TDVP in ITensorMPS.jl), although it performs strictly more arithmetic. The
extra arithmetic is visible where it can be isolated, as the 10\% in the early TDVP steps of Fig.~\ref{fig:tdvp}.
The gain has a different origin: the cost of managing a large live memory footprint.

\paragraph*{Garbage collection in Julia.} Julia's collector periodically marks all reachable objects, so the cost
of a full collection grows with the live heap. With every environment stored, the live heap at $D=758$ is
dominated by 72 environment blocks; with checkpointing ($k=8$) it holds about 25. In the TDVP runs of
Fig.~\ref{fig:tdvp} the time spent in garbage collection grows smoothly with $D$ in both cases and stays below
6\% of the step time with checkpointing. At $D=758$ the stored run crosses a threshold: 2079\,s of its 3716\,s
step are spent in the collector, against 62 and 82\,s in the two checkpointed runs. The checkpointed projector
calls the collector after dropping blocks. With that call removed, so that both runs follow Julia's default
policy, the checkpointed step takes 1724 and 1744\,s (54--58\,s in the collector) against 3405 and 3423\,s stored
(1550--1610\,s). The forced call thus accounts for about 8\% of the step and not for the difference. The peak
resident memory of the stored run is only 20\% higher (5.5 against 4.5\,GB) and far below the 200\,GB available,
which rules out swapping or memory exhaustion: the work of the collector scales with the number of live blocks.
Whether a given run crosses the threshold depends on the collector's heuristics, which explains why the same
stored step took 3716\,s in one run and more than 3870\,s in the other.

\paragraph*{Allocation and page faults in Python.} TeNPy allocates every environment as a fresh NumPy array.
Table~\ref{tab:tenpy} shows three alternated pairs on the dedicated Rome node and three runs on a second, Genoa,
node. The user CPU time, which measures the arithmetic, differs little between stored and checkpointed runs. On
the Rome node the system time and the number of minor page faults, which measure the kernel's work in mapping
fresh memory, fall several-fold with checkpointing, and the wall time falls by a factor of 1.09 (median of three
pairs; range 1.04--1.14). On the Genoa node the page faults fall 3.6-fold and the wall time by a factor of 1.15,
while the system time changes by about 12\%. The reduction in page faults is thus the common feature of the
faster runs. We attribute the gain to memory-mapping work that grows with the amount of memory a process keeps
mapped; its size depends on the machine.

\begin{table}[tbp]
\caption{TeNPy two-site DMRG, square $J_1$-$J_2$ YC6, $48\times6$, $D=1024$, $S^z$ conserved, 64 BLAS threads
(OpenBLAS), stored against checkpointed ($k=17$) environments, energies identical: wall time, system CPU time,
minor page faults and peak RAM. R: Rome, AMD EPYC 7542, whole node, pairs alternated. G: Genoa, AMD EPYC 9654,
128 of 192 cores allocated, two stored runs and one checkpointed run; the node was not reserved exclusively.}
\label{tab:tenpy}
\begin{ruledtabular}
\begin{tabular}{@{}lcccc@{}}
run & wall (s) & sys.\ (s) & faults ($10^6$) & GB\\
\hline
R1, stored & 12798 & 23383 & 836 & 13.5\\
R1, ckpt. & 11732 & 4035 & 216 & 3.44\\
R2, stored & 12983 & 25629 & 890 & 13.3\\
R2, ckpt. & 11381 & 5999 & 278 & 3.56\\
R3, stored & 12192 & 19438 & 733 & 13.5\\
R3, ckpt. & 11684 & 5845 & 275 & 3.56\\
G, stored & 2111 & 49256 & 580 & 13.3\\
G, stored & 2123 & 49964 & 578 & 13.3\\
G, ckpt. & 1831 & 43352 & 162 & 3.19\\
\end{tabular}
\end{ruledtabular}
\end{table}

These speedups are a side effect of the memory saving and not a property of the algorithm. In a code that
preallocates its environment storage, as our GPU code does through a memory pool, they are absent and only the
cost of recomputation remains. They are nevertheless what a user of these libraries will observe: in the CPU
benchmarks above, the memory saving usually came with no increase in wall time.

\section{Convergence at large bond dimension}
\label{sec:conv}

Completing sweeps at a larger $D$ is a test of memory, not of convergence. Whether the additional Schmidt states
improve the answer has to be checked separately, through the local optimisation, the truncation and the
finite-size effects~\cite{Stoudenmire2012,WhiteChernyshev2007}. As an illustration of calculations that
checkpointing brings within reach of one device, we followed the energies of two cylinders on a single H200 with
the Davidson-4 solver up to $D=3072$, beyond the stored limits of Table~\ref{tab:reach}.

\begin{figure}[tbp]
\includegraphics[width=\columnwidth]{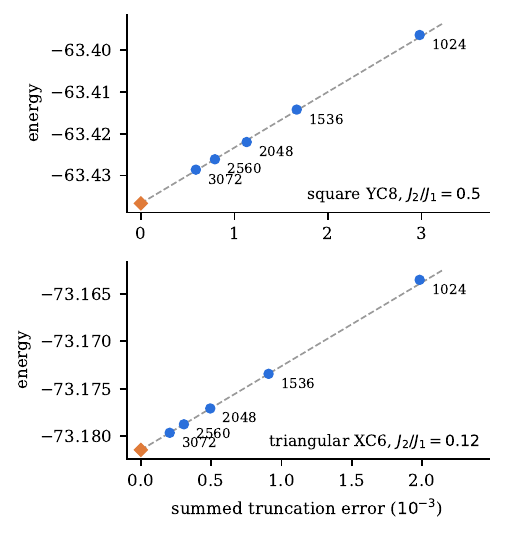}
\caption{Ground-state energy against summed truncation error at the last sweep of each bond dimension $D\ge1024$
(labels), with a linear fit through the four largest $D$ (dashed) and its intercept (diamond). All runs use
checkpointed environments on a single H200.}
\label{fig:extrap}
\end{figure}

For the square YC8 cylinder ($16\times8$, $J_2=0.5$) the energy at $D=3072$ is $-63.428586$, changing by
$7.7\times10^{-8}$ per sweep. Two further sweeps with tighter local solves (up to 24 matrix-vector products per
solve; largest residual $2.6\times10^{-4}$ after the first and $2.3\times10^{-8}$ after the second) shift the
energy by $+2.15\times10^{-6}$, to $-63.428584$. This change is much smaller than the extrapolation uncertainty
quoted below. The tighter local solves therefore support truncation-dominated error at this $D$, although they
do not certify convergence to the global ground state. A full energy variance was not evaluated at this size.
We estimate the remaining finite-$D$ bias by extrapolating in the truncation error~\cite{Hubig2018}. A linear
fit over the largest bond dimensions gives $E_0=-63.436\pm0.001$, where the uncertainty is the spread of fits
starting at $D=1024$, 1536 and 2048 (Fig.~\ref{fig:extrap}). This fit spread is not a rigorous error bound.
Subtracting an
$8\times8$ cylinder to remove the open ends gives a bulk energy of $-0.4985$ per site, consistent within about
$10^{-3}$ with the DMRG results of Gong \emph{et al.}~\cite{Gong2014}. For the triangular XC6 cylinder the energy converges to $-73.179646$ at $D=3072$, with an extrapolated
$E_0\approx-73.182$.

\section{Practical use}
\label{sec:practice}

Environment checkpointing is a small change with a predictable cost and no effect on results. It is most useful
where environments dominate memory: long systems, wide MPOs, large bond dimensions, and a GPU whose memory is the
limit. Its overhead, about one environment update per site and half-sweep for one level, cannot be hidden on a
device that is already busy and is largest with cheap local solvers.

\paragraph*{Estimating the benefit.} The gain follows from the composition of the peak. In our GPU code at
$D=1536$ (triangular XC6, $N=144$) the stored environments take 88.8\,GB, the working memory 18.4\,GB and the MPS
4.7\,GB. The working memory is equivalent to about 26 environment blocks, several times the $WD^2d^2$ of a single
two-site intermediate, and chunking brings it to about 19. Checkpointing pays once $N$ is well above this number,
and the saving grows with length, from 1.9$\times$ at $N=96$ to 3.8$\times$ at $N=384$ in peak memory
(Table~\ref{tab:length}). The same numbers bound what can be gained. With environments and working memory removed
altogether, the MPS alone would give an estimated gain in bond dimension of $\sqrt{111.9/4.7}\approx4.9$
relative to the unchunked stored baseline. For comparison, the earlier two-level search gave
$3712/1664\approx2.23$ relative to that baseline, including the saving from chunking. For short chains with
wide MPOs, as in ab initio chemistry with tens of orbitals, the gain is small
(Sec.~\ref{sec:cpu}). A single calibration sweep of a stored run suffices for an estimate: with $B$ and
$M_\mathrm{other}$ read from it, Eq.~\eqref{eq:mem} predicts the checkpointed peak to within 2\% near the optimal
spacing, and Eq.~\eqref{eq:time} the time to within 1\%. Extrapolation to a larger $D$ must allow for the growth
of the working memory and, in symmetric codes, of the symmetry blocks.

\paragraph*{Choosing the schedule.} The default in all our implementations is $k=\mathrm{round}(\sqrt{N})$. The
minimum of Eq.~\eqref{eq:mem} is flat and the measured optimum lies within one site of $\sqrt{N}$
(Table~\ref{tab:length}); a larger $k$ lengthens the rebuilt segment and a smaller one keeps more checkpoints,
with no gain in either direction. The two-level schedule should be reserved for cases in which one level does not
fit, since its recomputation grows linearly with $k_2$ [Eq.~\eqref{eq:time2}]. Chunking is best limited to about
four chunks, which give most of its saving for a few per cent in time at large $D$ (Sec.~\ref{sec:memory}).

\paragraph*{Combining memory-saving methods.} Checkpointing combines with symmetry-based block
sparsity~\cite{McCulloch2007,Singh2010}, mixed precision and offloading; bitwise comparisons then require the
precision and contraction settings to be held fixed. With offload, only the $N/k$ checkpoints need leave the
device, which lowers the host-memory requirement that limited offload in Sec.~\ref{sec:offload} from order $NB$
to order $(N/k)B$. Symmetric tensors benefit as well: with U(1) symmetry in ITensorMPS.jl the attainable bond
dimension on an H100 rose by a factor of 1.29.

\paragraph*{Scope.} The reconstruction argument holds whenever an algorithm consumes environments built from a
part of the state that has not changed. Beyond the DMRG and TDVP implementations tested here, this includes
excited-state optimisation~\cite{Stoudenmire2012}, finite-temperature evolution by purification or typical
states~\cite{Verstraete2004,White2009}, and variational MPS compression in the simulation of quantum circuits.
What matters is the contraction schedule and not the MPS form alone: gate-based evolution and MPO-based
propagation update the state differently~\cite{Vidal2004,Zaletel2015,Paeckel2019} and need not offer the same
saving. In out-of-core codes the schedule would reduce disk use instead of memory.

\paragraph*{Limitations.} The GPU measurements use one dense code without quantum-number symmetries; block
sparsity would shift the balance between environments and working memory. Bond-dimension limits are located on a
grid of 64 or 128, and the square-lattice values are lower bounds. The capped runs reproduce the memory of a
smaller card but not its bandwidth or speed. The combination of checkpointing with offload was not benchmarked.
The CPU speedups depend on the runtime and the machine (Sec.~\ref{sec:faster}). In quimb the moving environment
is built in full before the first block is dropped, so the steady-state memory is reduced but not the peak at
start-up. Where the working memory of a library dominates, as in dense ITensorMPS.jl or in quimb's
effective-Hamiltonian contraction at large sizes, that memory has to be reduced first.

\section{Conclusion}
\label{sec:conclusion}

The environments of a finite MPS sweep can be treated as a cache and not as part of the state: they are
determined by tensors that the sweep has not yet touched, and can be rebuilt exactly when needed. Keeping every
$\sqrt{N}$-th environment reduces their memory from $N$ to $2\sqrt{N}$ blocks at the cost of one environment
update per site and half-sweep, which adds 6--15\% to the sweep time for short local solves and less than 1\%
for converged ones. Earlier searches with checkpoints rebuilt at each half-sweep increased the attainable
bond dimension by factors of 1.77 for one level and 2.07 for two levels with eight chunks, relative to stored
baselines with matching chunking. With retained checkpoints at $D=1536$, the two-level schedule takes 1.42 times
the sweep time of the equally chunked stored baseline, or 1.79 times that of the unchunked stored baseline.
At the two-level limit measured with rebuilt checkpoints,
the working memory of the local update is the largest contribution to the peak. The method needs no host memory, complements
offloading, and is available for four widely used libraries. The bond dimension required on a cylinder grows
exponentially with its width, and after a quench exponentially with time. A factor of two in $D$ therefore
translates into a tighter extrapolation on the widest accessible cylinder and a fixed gain in reachable time, on
unchanged hardware.

\section*{Code and data availability}
The implementations, the GPU code, tests, run scripts and raw benchmark outputs are available at
\url{https://github.com/NihalShriv/envcheckpoint}. The version used for this paper,
v0.4.2, is archived at \href{https://doi.org/10.5281/zenodo.24193745}{10.5281/zenodo.24193745}~\cite{envcheckpoint};
\href{https://doi.org/10.5281/zenodo.23725022}{10.5281/zenodo.23725022} resolves all versions. The data behind the tables and figures and the figure-generation script are in the
repository's \texttt{results/} directory. The TeNPy implementation has been merged into the library
(\url{https://github.com/tenpy/tenpy/pull/713}). The ITensorMPS.jl contribution is under review
(\url{https://github.com/ITensor/ITensorMPS.jl/pull/228}).

\begin{acknowledgments}
The calculations were performed on the Sharanga high-performance computing facility of BITS Pilani,
Hyderabad Campus.

RB thanks support from an intramural grant CDRF C2/24/282 from BITS Pilani, and from ANRF India through ANRF/ARG/2025/009241/PS. RB acknowledges stimulating discussions with Indrakshi Raychowdhury on quantum algorithms, which motivated the present work.

ChatGPT Astra (OpenAI) was used to assist in framing the logical flow
of the manuscript, and Claude Code was used to develop the analysis code. The authors take responsibility for the final results and conclusions.
\end{acknowledgments}

\bibliography{refs}
\end{document}